\documentclass[12pt]{elsarticle}
\usepackage{graphicx}
\usepackage{amssymb}    
\usepackage{epstopdf}
\usepackage{comment}
\usepackage{color}
\usepackage{hyperref}
\usepackage{algorithm,algorithmic}
\usepackage[final]{pdfpages}
\usepackage{frcursive}
\usepackage{bbm}
\usepackage{mathtools}
\usepackage{tabularx}
\usepackage{algorithm,algorithmic}

\usepackage{yfonts}
\usepackage{lscape}
\usepackage{epsfig}
\usepackage{fullpage}
\usepackage[titletoc]{appendix}
\usepackage{fancyhdr}
\usepackage{amsmath,amssymb,xspace}

\usepackage{tabularx}
\usepackage{mathrsfs}

\newfont{\bbb}{msbm10 scaled\magstep1}

\newtheorem{rem}{Remark}[section]

\let \leq \leqslant
\let \geq \geqslant

\let \epsilon \varepsilon

{ \par \medskip \par
  \noindent \textit{\textbf{Demonstration\/}} : }{\null \hfill $\Box$ \par }
 
\newcommand{\R} {\ensuremath{\mathbb{R}}}
{

\newcommand{\N} {\ensuremath{\mathbb{N}}}
\newcommand{\C} {\ensuremath{\mathbb{C}}}

\newcommand{\W}{\mathscr{W}}  

\makeatletter
\newcommand{\doublewidetilde}[1]{{%
  \mathpalette\double@widetilde{#1}%
}}
\newcommand{\double@widetilde}[2]{%
  \sbox\z@{$\m@th#1\widetilde{#2}$}%
  \ht\z@=.9\ht\z@
  \widetilde{\box\z@}%
}

\begin{document}

\begin{frontmatter}


\title{Computation of the Time-Dependent Dirac Equation with Physics-Informed Neural Networks}

\author[carl,crm]{Emmanuel LORIN}
\ead{elorin@math.carleton.ca}
\author[ucsb]{Xu YANG}
\ead{xuyang@math.ucsb.edu}

\address[carl]{School of Mathematics and Statistics, Carleton University, Ottawa, Canada, K1S 5B6}
\address[crm]{Centre de Recherches Math\'{e}matiques, Universit\'{e} de Montr\'{e}al, Montr\'{e}al, Canada, H3T~1J4}
\address[ucsb]{Department of Mathematics, University of California, Santa Barbara, CA 93106, USA}

\begin{abstract}
We propose to compute the time-dependent Dirac equation using physics-informed neural networks (PINNs), a new powerful tool in scientific machine learning avoiding the use of approximate derivatives of differential operators. PINNs search solutions in the form of parameterized (deep) neural networks, whose derivatives (in time and space) are performed by automatic differentiation. The computational cost comes from the need to solve high-dimensional optimization problems using stochastic gradient methods and train the network with a large number of points. Specifically, we derive PINNs-based algorithms and present some key fundamental properties of these algorithms when applied to the Dirac equations in different physical frameworks. 
	
\end{abstract}



\end{frontmatter}

\section{Introduction}
In this paper, we are interested in computing the relativistic two-dimensional time-dependent Dirac equation (TDDE) using physics-informed neural network (PINN) algorithms \cite{pinns,pinns2,pinns3,pinns4}. PINN is a new method for solving partial differential equations (PDEs) using neural networks (NNs), which takes its origin in the celebrated work of Lagaris \cite{lagaris}. Generally speaking, the strength of the PINN approach is its impressive flexibility, particularly regarding the dimensionality of the PDE under consideration, and its mathematical structure, at least for linear PDEs. However, solutions to nonlinear PDE systems may require much higher computational resources. We will see, for instance, that when solving the Dirac equation in flat or curved spaces, the structure of the PINN algorithm is identical, unlike standard PDE solvers, which require more technical attention in curved space (regarding the numerical stability, or the order of convergence). More specifically, a loss function is constructed as the norm of the PDE operator under consideration applied to the searched neural network. This brings the following advantages: i) PINN does not require the discretization of any differential operator; ii) is embarrassingly parallel; and iii) is easy to implement (at least using neural network libraries). In particular, the derivatives of neural networks are performed exactly using automatic differentiation. The PINN algorithm applied to the space-time differential operator is not based on an evolution procedure but on the minimization of nonlinear functions thanks to stochastic gradient methods. The counterpart is that the minimization algorithm (typically stochastic gradient method \cite{sgm1,sgm2}) usually does not converge to a global minimum. Hence, we have to make sure that the computed solution is close enough to the exact solution by choosing the optimized algorithm properly. Additional constraints (such as mass conservation) may be used to guide the minimization algorithm. Unlike, for instance, recent by Grobe et al. \cite{grobe2}, the PINN algorithm presented below (at least by default) has no requirement on the structure of the solution to the Dirac equation. Here we propose a pedagogical presentation of the algorithms, particularly without prior scientific machine learning knowledge, starting with introducing the TDDE under consideration. The intent of this paper is then to give some relevant information (pros and cons) on the use of PINN algorithms for solving the TDDE.

\subsection{Time-dependent Dirac equation}
We consider the following two-dimensional time-dependent Dirac equation on $\Omega\times [0,T]$ \cite{thaller} ($\Omega \subseteq \R^2$)
\begin{eqnarray*}
{\tt i}\partial_t\psi =  -\big\{\sigma_x({\tt i}c\partial_x+eA_x({\boldsymbol x},t)) + \sigma_y({\tt i}c\partial_y +  eA_y({\boldsymbol x},t)) - mc^2 \sigma_z -  V({\boldsymbol x})I_2 - A_0({\boldsymbol x},t)I_2\big\}\psi \, ,
\end{eqnarray*}
and initial condition $\psi({\boldsymbol x},0)=\phi_0({\boldsymbol x}) \in L^2(\Omega;\C^2)$, where $c$ denotes the speed of light, $m$ is the mass of the electron, and where Pauli's matrices are defined by
\begin{eqnarray}
\sigma_{x} = 
\left(
\begin{array}{cc}
0 & 1 \\ 1 & 0  
\end{array}
\right)
\;\; \mbox{,} \;\;
\sigma_{y} = 
\left(
\begin{array}{cc}
0 & -{\tt i} \\ {\tt i} & 0 
\end{array}
\right)
\;\; \mbox{and} \;\;
\sigma_{z} = 
\left(
\begin{array}{cc}
1 & 0 \\ 0 & -1 
\end{array}
\right) \, .
\end{eqnarray}
For $\Omega$ bounded, we impose null Dirichlet boundary condition on $\Omega$. Function $V$ represents an interaction potential, and $(A_x,A_y)$, $A_0$ denote an external space-time dependent electromagnetic field which is assumed to be given. In the following, we will also denote $\widetilde{\sigma}_y:=-{\tt i}\sigma_y$. In the framework of this paper, the Dirac equation typically models  relativistic fermion \cite{thaller,Itzykson:1980rh} possibly interacting with an external electromagnetic fields. There exists an important literature on computational methods for solving this problem; see for instance \cite{cpc2017,keitel,keitel4,Fillion-Gourdeau2016122,0022-3700-19-20-003,grobe}.

Specifically, we are interested in the proof of feasibility, and relevance of the PINN method in the framework of relativistic quantum physics. In particular, the focus will be put on the behavior of this methodology regarding some fundamental properties from the computational and physical points of view: scaling when $c$ increasing, numerical dispersion (Fermion doubling \cite{potz2}), and preservation of the $L^2$-norm.

Using the PINN algorithm for the TDDE allows to avoid the discretization in space$\&$time derivatives with a loss function which has a {\it natural} structure.  In particular, we can avoid the usual stability constraints on the time step, as it is no more necessary to discretize in time \cite{bao,bao2,bao3,cpc2012,cpc2017}. The presence of the external electromagnetic field would for instance complexify the stable and accurate discretization of the Dirac equation with standard approximate PDE solvers \cite{cpc2017,cpc2012}. Hereafter, for computational convenience, we rewrite the Dirac equation in the form of a {\it real} $4$-equation system
\begin{eqnarray}\label{HD}
\left.
\begin{array}{lcl}
\mathcal{H}_D({\boldsymbol x},t)\Phi & := & \partial_t \Phi + H_x \partial_x \Phi + H_y \partial_y \Phi + C({\boldsymbol x},t) \Phi \, ,
\end{array}
\right.
\end{eqnarray}
where $\Phi=(\psi_R,\psi_I) \in L^2(\Omega\times [0,T];\R^4)$ the real and imaginary parts of the Dirac equation solution $\psi \in L^2(\Omega\times [0,T];\C^2)$, and where
\begin{eqnarray*}
H_x = c\left(
\begin{array}{cc}
\sigma_x & 0_2 \\
0_2 & \sigma_x 
\end{array} 
\right), \, \, \, 
H_y = c\left(
\begin{array}{cc}
0_2 & -\widetilde{\sigma}_y \\
\widetilde{\sigma}_y & 0_2 
\end{array}
\right) \, ,
\end{eqnarray*}
and where $C({\boldsymbol x},t)$ is given by
\begin{eqnarray*}
\left(
\begin{array}{cc}
0_2 & -V({\boldsymbol x})I_2 - mc^2 \sigma_z +eA_x({\boldsymbol x},t)\sigma_x+eA_y({\boldsymbol x},t)\widetilde{\sigma_y} \\
V({\boldsymbol x})I_2+mc^2\sigma_z - eA_x({\boldsymbol x},t)\sigma_x - eA_y({\boldsymbol x},t)\widetilde{\sigma_y} & 0_2
\end{array}
\right).
\end{eqnarray*}
We denote by $\Phi_0=(\phi_0,{\boldsymbol 0})^T\in L^2(\Omega;\R^4)$ the initial condition. Let us recall that the Dirac equation is a Hermitian linear first-order hyperbolic system \cite{11}. In the following, we will denote $H_D({\boldsymbol x},t):=H_x\partial_x + H_y\partial_y+C({\boldsymbol x},t)$.  Although $C$ is an antisymmetric matrix, the proposed algorithm will compute the ``correct'' real solution, see \cite{deepXDE}.
\subsection{Neural networks}
We here recall the basics of neural network-based algorithms for solving PDE. The principle consists in searching for the solution to the PDE under consideration, in the form of a parameterized deep neural network and to search for the corresponding parameters by minimizing a loss function.  Denoting the neural network $N({\boldsymbol w},{\boldsymbol x})$ with ${\boldsymbol x}=(x_1,\cdots,x_d) \in \Omega \subseteq \R^d$ and  by ${\boldsymbol w}$ the unknown parameters. Neural networks usually read (for 1 hidden layer, machine learning)
\begin{eqnarray}\label{NN0}
\left.
\begin{array}{lcl}
N({\boldsymbol w},{\boldsymbol x}) & = & \sum_{i=1}^Hv_{i}\sigma_{i}\big(\sum_{j=1}^dw_{ij}x_j+u_i\big) \, ,
\end{array}
\right.
\end{eqnarray}
where $\{\sigma_i\}_{1\leq i \leq H}$ are the sigmoid transfer functions, and $H$ is the number of sigmoid units (Neurons), $\{w_{ij}\}_{ij}$ are the weights and $\{u_i\}_i$ the bias. When considering several hidden layers (deep learning), we have then to compose functions of the form \eqref{NN0}, \cite{despres}. That is,
\begin{eqnarray*}
N & = & \mathcal{N}_p \circ  \mathcal{N}_{p-1}\circ \cdots \mathcal{N}_1\circ  \mathcal{N}_0 \, ,
\end{eqnarray*}
where for $0\leq r\leq p$, $\mathcal{N}_r$ is defined from $\R^{a_r}$ (with $a_{r} \in \N$) to $\R^{a_{r+1}}$ by $\sigma_r(W_rX_r+b_r)$, $\sigma_r$ is an activation function, $X_r\in \R^{a_r}$ and where $(a_0,\cdots,a_{p+1})$ where $p+1$ layers are considered. The layer $r=0$ is the input layer and $r=p+1$ is the output layer, such that $a_0=a_{p+1}=m$. We refer to \cite{NNbook} details about neural networks. 
\\
In this paper, we propose to adapt the PINN algorithm for different versions of the Dirac equation in different frameworks, such as relativistic quantum physics, and strained graphene. Beyond the simple derivation of the algorithm and its computational complexity, we are also interested in some fundamental properties of this solver. In particular, some crucial physical constraints such $L^2$-norm conservation of the wavefunction or fermion-doubling \cite{cpc2012}. 
\subsection{Organization}
This paper is organized as follows. In Section \ref{sec:pinn}, we introduce the PINN algorithm in the framework of the TDDE, and derive some basic computational and mathematical properties of the method. Section \ref{sec:numerics} is dedicated to some illustrating numerical experiments. Some concluding remarks are finally proposed in Section \ref{sec:conclusion}.
\section{PINN algorithm for the Dirac equation}\label{sec:pinn}
In this section, we present the principle of the PINN algorithm applied to the TDDE. 
\subsection{Presentation of the PINN algorthim}
We search for an approximate solution to this evolution PDE in the form of a space-time dependent vector valued deep neural network $(N_1,N_2,N_3,N_4)^T={\boldsymbol N} \in L^2 (\W\times \Omega\times[0,T];\R^4)$, where $W \subset \R^p$ denotes the search space. In order to include the initial condition, we can search for NN in the form
\begin{eqnarray*}
{\boldsymbol N} \longleftarrow {\boldsymbol \Phi}_0 + f(t) {\boldsymbol N} \, , \text{with}\quad f(0)=0.
\end{eqnarray*}
For example, one may  take $f(t)=t$ as in \cite{lagaris}. Alternatively, it is possible to pick other functions allowing to consider larger times such as $f(t)=(1-e^{-\gamma t})/\gamma$ for some free parameters $\gamma>0$. In particular, when $t$ goes to infinity, the contribution of the initial data in the loss function does not become negligible; allowing to avoid convergence of the PINN algorithm to a null solution. In addition for long physical times, it may be necessary to impose a $L^2$-norm constraint: $\|{\boldsymbol N}(\cdot,t)\|_{L^2(\Omega)}=  \|{\boldsymbol N}_0\|_{L^2(\Omega)}=1$. In the latter case, the PINN algorithm hence consists in computing for some free parameter $\lambda\geq 0$
\begin{eqnarray*}
\overline{{\boldsymbol p}} & = & \textrm{argmin}_{{\boldsymbol p} \in W} \big\{\big\|\mathcal{H}_D(\cdot,\cdot){\boldsymbol N}({\boldsymbol p},\cdot,\cdot)\big\|^2_{L^2(\Omega\times(0,T))} + \lambda \big\| |{\boldsymbol N}({\boldsymbol p},\cdot,\cdot)|_{\ell^2}-1\|^2_{L^2(\Omega\times (0,T))} \big\} \, ,
\end{eqnarray*}
where $\mathcal{H}_D$ is defined in \eqref{HD}. Additional constraints can naturally easily be added to force some conditions. A fundamental element of the algorithm is the automatic differentiation of the neural network which allows to perform exactly the derivative of NN. The counterpart is the regularity of the overall solution is naturally imposed by the regularity of the neural network. With such minimization, we ensure that the searched function is indeed an approximate solution to the Dirac equation, and that it approximately satisfies the $L^2$-norm conservation. An alternative approach based on time-domain decomposition is also proposed in Section \ref{sec:numerics}. In this paper, we will focus on the computation of the density
\begin{eqnarray}\label{density}
\rho_N({\boldsymbol x},t) & = & \sum_{i=1}^4|N_i(\overline{{\boldsymbol p}},{\boldsymbol x},t)|^2 \, .
\end{eqnarray}
It is also interesting to notice that in the above algorithm, the searched neural network is a space- and time-dependent function. In other words, the initial condition is considered as a Dirichlet boundary condition in space-time domain. As a consequence, we get an approximate solution of the TDDE at any time and any point in space. The inclusion of more complex boundary conditions will require to add some constraints to the loss function.
\\
Rather than or in addition, to the constraint on the loss function forcing the $L^2$-norm of the solution to $1$, we propose a natural  alternative approach, for long physical times. This simple idea consists in decomposing the physical time as follows $[0,T] = \cup_{i=0}^N[t_i,t_{i+1}]$ with $t_0=0$ and $t_{N+1}=T$ and with $\Delta t = t_{i+1}-t_i$ small enough to unsure the $L^2$-norm preservation during each ``small'' space-time subdomain $\Lambda_{i}:=\Omega\times [t_i,t_{i+1}]$. 
On each of these space-time $\Lambda_{i}^2$ intervals, we apply the same strategy$/$method as above. However in this case, most of the training is performed in the spatial directions. Assuming $\Phi_i$ given (and computed on $\Lambda_{{i-1}}$), we solve on $\Lambda_{i}$
\begin{eqnarray}\label{DDT1}
\left.
\begin{array}{lcl}
\partial_t \Phi_{i+1} & =& \mathcal{H}_{D}({\boldsymbol x},t)\Phi_{i+1} \\
\Phi_{i+1}(\cdot,t_{i})& = & \Phi_i(\cdot,t_i) \, .
\end{array}
\right.
\end{eqnarray}
This system is solved by minimizing on each interval $(t_i,t_{i+1})$
\begin{eqnarray}\label{DDT2}
\left.
\begin{array}{lcl}
\overline{{\boldsymbol p}}_{i+1} & = & \textrm{argmin}_{{\boldsymbol p} \in W} \big\{\big\|\mathcal{H}_D(\cdot,\cdot){\boldsymbol N}_{i+1}({\boldsymbol p},\cdot,\cdot)\big\|^2_{L^2(\Lambda_{i})}\\
& &  + \lambda \big\| |{\boldsymbol N}_{i+1}({\boldsymbol p},\cdot,\cdot)|_{\ell^2}-1\|^2_{L^2(\Lambda_{i})} \big\} \, ,
\end{array}
\right.
\end{eqnarray}
where ${\boldsymbol N}_{i+1}(\overline{{\boldsymbol p}}_{i+1},\cdot,\cdot)$ is an neural network approximating the solution $\Phi_{i+1}$.  Notice that this decomposition in time, theoretically does not change the overall solution if it were performed exactly by the PINN algorithm. That is $\Phi_{N+1}=\Phi$ where $\Phi$ is the exact solution to \eqref{HD}. Hence, we just need to make sure that, at least on each interval the $L^2$-norm of $\Phi_i$ is preserved. The drawback of this time-domain decomposition is the loss of the embarassingly parallelization. 
\begin{rem}\label{rem1}
Alternatively to the approach proposed above, it is also possible to apply the PINN algorithm in space only. In the latter case, we however need to approximate the time derivative in $\mathcal{H}_D$. In this goal, we introduce discrete times $t_0<t_1<\cdots<t_n< t_{n+1}<\cdots$. Assuming that ${\boldsymbol N}^n(\overline{{\boldsymbol p}}^n,{\boldsymbol x})$ is a known function (computed at previous time iteration), we write
\begin{eqnarray*}
{\boldsymbol N}^{n+1}({\boldsymbol p},{\boldsymbol x}) & =& {\boldsymbol N}^n(\overline{{\boldsymbol p}}^n,{\boldsymbol x}) + \int_{t_n}^{t_{n+1}}H_D({\boldsymbol x},s){\boldsymbol N}({\boldsymbol p},{\boldsymbol x},s)ds \, .
\end{eqnarray*}
It is then necessary to approximate the time-integral. We propose to use a trapezoidal quadrature leading to
\begin{eqnarray*}
\Big(I-\cfrac{t_{n+1}-t_n}{2}H_D({\boldsymbol x},t_{n+1})\Big){\boldsymbol N}^{n+1}({\boldsymbol p},{\boldsymbol x}) & =& \Big(I+\cfrac{t_{n+1}-t_n}{2}H_D({\boldsymbol x},t_{n})\Big){\boldsymbol N}^n(\overline{{\boldsymbol p}}^n,{\boldsymbol x})\, .
\end{eqnarray*}
Hence the PINN-in space algorithm consists in minimizing the following loss function
\begin{eqnarray*}
\mathcal{L}^{n+1}({\boldsymbol p}) = \Big\|\Big(I-\cfrac{t_{n+1}-t_n}{2}H_D(\cdot,t_{n+1})\Big){\boldsymbol N}^{n+1}({\boldsymbol p},\cdot) - \Big(I+\cfrac{t_{n+1}-t_n}{2}H_D(\cdot,t_{n})\Big){\boldsymbol N}^n(\overline{{\boldsymbol p}}^n,\cdot)\Big\|_{L^2(\Omega)} \, .
\end{eqnarray*}
Interestingly, and unlike standard PDE implicit solvers, the solution is not updated at time $t_{n+1}$ by directly solving a linear system, but by minimizing the loss function. The approximate solution at time $t_{n+1}$ is given by:
\begin{eqnarray*}
{\boldsymbol N}^{n+1}\big(\textrm{argmin}_{{\boldsymbol p} \in W}\mathcal{L}^{n+1}({\boldsymbol p}),{\boldsymbol x}\big) \, .
\end{eqnarray*}
Notice that if we assume that  $V$ and ${\boldsymbol A},A_0$ are null, $H_D$ is a constant operator and by Parseval's identity and hyperbolicity of the Dirac Hamiltonian, we have
\begin{eqnarray*}
\left.
\begin{array}{lcl}
\|{\boldsymbol N}^{n+1}(\overline{{\boldsymbol p}}^{n+1},\cdot)\|_{L^2(\R^4)} & \approx& \Big\|\Big(I-\cfrac{t_{n+1}-t_n}{2}H_D\Big)^{-1} \Big(I+\cfrac{t_{n+1}-t_n}{2}H_D\Big){\boldsymbol N}^n(\overline{{\boldsymbol p}}^n,\cdot)\Big\|_{L^2(\R^4)} 
\end{array}
\right.
\end{eqnarray*}
The latter can potentially be relaxed the requirement on the conservation of the $L^2$-norm, if the initial NN-based density has a $L^2$-norm equal to $1$. The main interest compared to using a standard PDE solver, is indeed the fact that we avoid the computation of a very high dimensional linear system if we search for a very accurate$/$precise solution (corresponding to a fine grid in space). 
\end{rem}
The PINN algorithm requires the computation of the NN parameters obtained thanks to the minimization of the loss function ({\it training process}):
\begin{itemize}
\item We first compute explicitly the loss functions at some {\it training points} $\{({\boldsymbol x}_j,t_n)\}_{(j,n) \in \N^{(i)}_{\boldsymbol x}\times \N_t}$ (resp.$\{({\boldsymbol x}_j,t_n)\}_{(j,n) \in \N^{(e)}_{\boldsymbol x}\times N_t}$) in the interior (resp. exterior) of the space-time domain. We denote by $\mathcal{N}_{{\boldsymbol x}}^{(i)}$ (resp. $\mathcal{N}_{{\boldsymbol x}}^{(e)}$ and $\mathcal{N}_{t}$) the number of elements in the set $\N_{{\boldsymbol x}}^{(i)}$ (resp. $\N_{{\boldsymbol x}}^{(e)}$ and $\N_t$). For instance, with Dirichlet boundary conditions the loss function (without the normalization constraint) is numerically minimized in the form
\begin{eqnarray*}
\left.
\begin{array}{lcl}
\mathcal{L}({\boldsymbol p})& = & \cfrac{1}{\mathcal{N}_{{\boldsymbol x}}\mathcal{N}_t}\sum_{(j,n)  \in  \N^{(i)}_{\boldsymbol x}\times \N_t} \big|\mathcal{H}_D({\boldsymbol x}_j,t_n){\boldsymbol N} ({\boldsymbol p},{\boldsymbol x}_j,t_n)\, \big|_{2}^2 \\
& & + \cfrac{1}{\mathcal{N}^{(e)}_{{\boldsymbol x}}\mathcal{N}_t}\sum_{(j,n)  \in  \N^{(e)}_{\boldsymbol x}\times \N_t} \big|{\boldsymbol N} ({\boldsymbol p},{\boldsymbol x}_j,t_n)\, \big|_{2}^2 \, .
\end{array}
\right.
\end{eqnarray*}
This process is very computationally complex, although its parallelization is straightforward .
\item Minimization of the loss function is performed using stochastic gradient descent (SGD) method \cite{sgm0,sgm1,sgm2}. The interest of the SGD is to deal with (several) lower dimensional optimization problems allowing to avoid the use of a deterministic gradient method on a high dimensional search space. The latter requires to fix some parameters such as the size of the epochs (corresponding to the number of iteration in the model training) and a learning rate.

\item Standard PDE solvers require the use of a time step less than $1/mc^2$ in order to precisely approximate the mass-term \cite{cpc2017} and the phenomenon called {\it zeitterbewegung} \cite{thaller}. Practically, this often limits  numerical simulations for TDDE to very short physical times. PINN algorithm can not circumvent this issue as the TDDE solution, even if it is not constructed on an evolution process.  More specifically, at a final time $T$, the number of training points in the $t$-direction should scale in $Tmc^2$.  Hence, there is a need for dealing with a very high dimensional minimization problem when $c$ and $T$ are large. Assuming that the solution does not contain high wavenumbers (space frequency), and  in order to avoid an overfitting in space, however we can take $\N_{\boldsymbol x}^{(i,e)}$ small. Similarly, in order to avoid underfitting in time, we need to take $\N_t^{(i,e)}$ large enough.  
\item Practically, we compute the solution to the Dirac equation on a bounded spatial subset $\Omega=(-L,L)^2$ of $\R^2$, and we impose Dirichlet boundary conditions at $(\pm L,y)$ and $(x,\pm L)$ which can easily be imposed through additional constraints on the loss function.
\end{itemize}
In order to improve the accuracy or precision of the NN-based solver, it is naturally possible to directly impose some constraints (symmetry, ) on the structure of the functions involved in the neural networks rather than imposing those constraints through Lagrange multipliers which would complexify the optimization.
\subsection{Computational complexity}\label{subsec:CC}
Let us now discuss the computational complexity of the PINN approach (in its most simple form) versus a standard implicit real space (finite volume, finite difference, finite element) PDE solver on a $N_{\boldsymbol x}$ spatial grid points with $N_t$ time iterations. The overall computational complexity of an implicit real space solver is $O(N_tN_{\boldsymbol x}^{\ell})$, where $1<\ell \leq 3$ depends on the sparsity of linear systems involved in the spatial discretization. Let us now discuss the complexity of the PINN algorithm in its most direct and simple setting. We denote by $\mathcal{N}_{\boldsymbol x}$ (resp. $\mathcal{N}_t$) the total number of interior and boundary spatial (resp. temporal) training points, and by $\mathcal{N}_{\boldsymbol p}$ the dimension of the search space; the latter depends on the depth of the NN (number of hidden layers and neurons). Naturally, as we search for space-time dependent NN, the dimension of the search space must be increased accordingly, compared to the space-only NN, as described in Remark \ref{rem1}. However practically $\mathcal{N}_{\boldsymbol p} \ll \mathcal{N}_{\boldsymbol x}+\mathcal{N}_t$.
\begin{itemize} 
\item The construction of the loss function does not require the computation of linear systems and is hence a (embarrassingly parallel) linear process requiring {\it only} $O(\mathcal{N}_{\boldsymbol x}\mathcal{N}_t)$ operations. 
\item The minimization of the loss function is usually based on a (stochastic) gradient method.  The latter requires the solution to linear systems that is $O(\mathcal{N}_{{\boldsymbol p}}^{m})$ operations for $1<m\leq 3$, depending on the sparsity and structure of $\nabla_{\boldsymbol p} \mathcal{L}$. In order to reduce the computational complexity, this step is done using a stochastic gradient method allowing to consider lower dimensional intermediate minimization problems \cite{sgm0,sgm1,sgm2}.
\end{itemize}
Hence overall and its in most simple setting, the PINN algorithm has a complexity of $O(\mathcal{N}_{{\boldsymbol p}}^{m}) + O(\mathcal{N}_{\boldsymbol x}\mathcal{N}_t)$. For very large computational times and compared to PDE solvers using fine grids and high order approximations (corresponding to $N_{\boldsymbol x}\gg 1$ and $\ell$ close to $3$), the PINN approach can become competitive. The analysis of accuracy of PINN algorithms is not based on the order of approximation of the space$/$time derivatives in the Dirac equation and$/$or on the choice of basis functions. Increasing the precision in the PINN framework has a different meaning and can be performed in different ways. The most natural way to increase the precision of the PINN solver is to increase the number of hidden layers and neurons, which is basically equivalent to increase the size of the function space in which we search for the solution. Alternatively or in addition, the optimization solver can be refined$/$improved by increasing the number of training points. The latter procedure usually provides a lower local minimum due to the complex structure of the loss function.  The latter is naturally one of the main weaknesses of the PINN algorithm.
\begin{rem}{\bf Extension to TDDE in curved space.} In this remark, we discuss the application of the PINN to the TDDE in curved space. The latter is in particular used to model strained graphene surfaces \cite{graphene1,graphene2,pre,pre2,chai}. In the case of stationary surfaces, the TDDE has space-dependent Pauli$/$Dirac matrices. The corresponding Dirac equation models non-relativistic electrons in the vicinity of Dirac points (low energy limit). Specifically for a general smooth two-dimensional surface, and neglecting the non-diagonal terms of the metric tensor associated to the surface \cite{pre}, the systems reads
\begin{eqnarray*}
\mathcal{H}_D({\boldsymbol x})\psi({\boldsymbol x},t) &= 0\, ,
\end{eqnarray*}
where
\begin{eqnarray*}
\mathcal{H}_D({\boldsymbol x}) &= {\tt i}\partial_t + {\tt i}\cfrac{v_{F}}{\sqrt{\rho(\boldsymbol{x})}}\biggl\{\sigma_x   \left[\partial_{x} + \Omega_{x}({\boldsymbol x})\right] + \sigma_y   \left[\partial_{v} + \Omega_{v}({\boldsymbol x}) \right] \biggr\} \, ,
\end{eqnarray*}
with affine spin connection (pseudomagnetic field)
\begin{eqnarray*}
\Omega_{x}(\boldsymbol{x}) = \cfrac{{\tt i}}{4} \cfrac{\rho_{v}(\boldsymbol{x})}{|\rho(\boldsymbol{x})|} \sigma_z, \, \, \Omega_{v}(\boldsymbol{x}) = -\cfrac{{\tt i}}{4} \cfrac{\rho_{x}(\boldsymbol{x})}{|\rho(\boldsymbol{x})|} \sigma_z  \, ,
\end{eqnarray*}
where $\rho$ is a smooth surface-dependent function, and $v_F$ is the Fermi velocity for unstrained graphene surface. We refer to \cite{pre} for details. Interestingly, the PINN again simply reads
\begin{eqnarray*}
\overline{{\boldsymbol p}} & = & \textrm{argmin}_{{\boldsymbol p} \in W} \big\|\mathcal{H}_D(\cdot){\boldsymbol N}({\boldsymbol p},\cdot,\cdot) \big\|^2_{L^2(\Omega\times(0,T))}\, .
\end{eqnarray*}
The presence of non-constant coefficients will only complexify a bit the structure of the loss function.
\end{rem}
\section{Numerical experiments}\label{sec:numerics}
In this section, we propose several numerical experiments to illustrate the above method applied to TDDE. The latter will be solved on a bounded domain with null Dirichlet boundary conditions. The implementation was performed using {\tt DeepXDE} \cite{deepXDE} and {\tt tensorflow} \cite{tensorflow2015-whitepaper} which are very flexible and easy-to-use libraries. In the tests below, we have taken $\lambda=0$, $c=1$ and we use $\tanh$ as transfer function. The corresponding codes will be made available on {\tt github}.\\
\\
\subsection{Numerical properties}\label{sec:numerics0}
\begin{figure}[hbt!]
\begin{center}
\includegraphics[height=6cm,keepaspectratio]{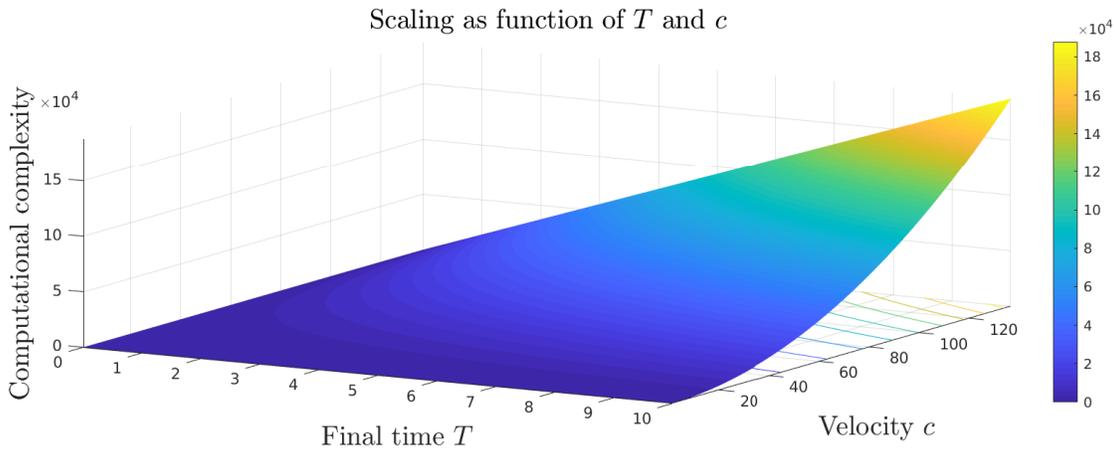}
\end{center}
\caption{Scaling of the computational complexity a function of $T$ and $c$.}
\label{fig000}
\end{figure}

\begin{figure}[hbt!]
\begin{center}
\includegraphics[height=10cm,keepaspectratio]{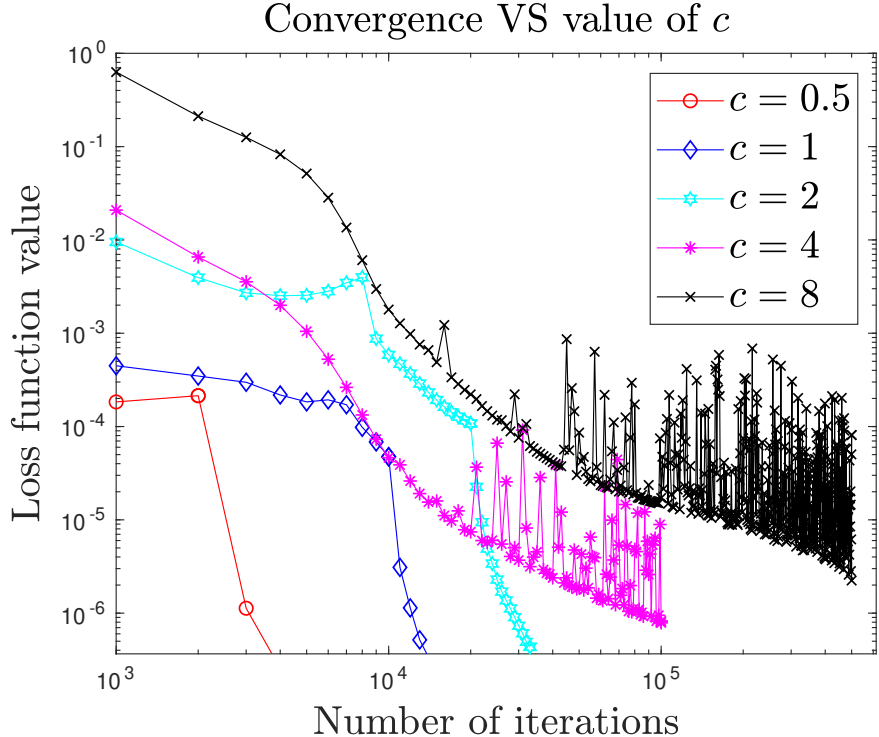}
\end{center}
\caption{{\bf Experiment 1.} Loss function convergence for $c=2^{i}$, $i=-1,\cdots,3$, with the number of training points given in Table \ref{tab1}.}
\label{fig0000}
\end{figure}

We are first interested in this subsection in some important numerical and physical properties of the PINN solver for the Dirac equation, in particular in the framework of relativistic quantum physics. More specifically, we discuss several important questions.\\
\noindent{\bf Scaling of the PINN solver from the computational point of view for increasing values of $c$}. As discussed above the typical time scale to capture the zitterbewegung requires time resolution of the order $1/mc^2$; at final time $T$, the number of training points in the $t$-direction should scale in $Tmc^2$. As the PINN-algorithm is linear as a function the number of training points (in any space- or time-direction), see Section \ref{subsec:CC}, the complexity of the PINN-algorithm should increase linearly in $T$ and quadratically in $c$, see Fig. \ref{fig000} for $(c,T)\in[1,137]\times[0,10]$.  The analogue requirement for standard numerical TDDE solvers, is the need  to take time steps smaller than $1/mc^2$.\\
\noindent{\bf Experiment 1.}
It is not that simple to numerically reproduce this behavior due to the large number of the parameters in the PINN algorithm. In particular, changing the value of the $c$, that is changing the value of a physical variables in the Dirac equation, we do not expect similar solutions. We then propose to proceed as follows. 
\begin{itemize}
\item The following physical data: $V$ ($V({\boldsymbol x})=-1/\sqrt{\|{\boldsymbol x}\|^2+1}$) , $m=1$, space-time domain $(-8,8)\times (0,0.25)$; and numerical data: number of hidden layers ($5$) and neurons ($30$), learning rate $10^{-4}$, will remain unchanged.
\item The following numerical parameters are changed: number of training points and number of training iterations (epoch).
\end{itemize}
We consider below $c$ equal to $2^{i}$, $i=-1,0,1,2,3$. We report in the Fig. \ref{fig0000} the loss functions for the different values of $c$ to show the convergence.

\begin{figure}[hbt!]
\begin{center}
\includegraphics[height=8cm,keepaspectratio]{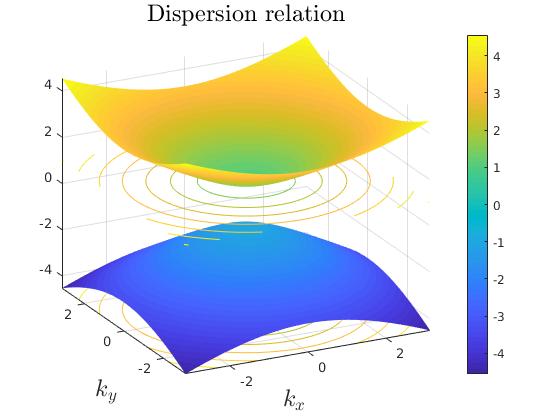}
\end{center}
\caption{Exact dispersion relation: graph of $\mu_{1,2}$ in \eqref{mu}.}
\label{fig00}
\end{figure}

\begin{table}[t!]
	\centering
	\caption{Number of training points and CPU as function of  $c$}
	\begin{tabular}{lll}
		\hline \hline
		Value of $c$ & Number of training points & CPU time (seconds)\\
		\hline
		$c=0.5$ & $5\times 10^3$ & $72.2$  \\
		$c=1$ & $7.5\times 10^3$ & $107.4$\\
		$c=2$ & $5\times 10^4$ & $242.6$ \\
		$c=4$ & $5\times 10^6$ & $1655.7$ \\
		$c=8$ &  $10^6$ & $6443.8$ \\
		\hline \hline
	\end{tabular}
	\label{tab1}
\end{table}
This experiment does not allow to rigorously reproduce the theoretical quadratic scaling CPU vs value of $c$. However it illustrates the rapid nonlinear increase of  the computational complexity when linearly increasing the value of $c$ due to the need to increase the number of training points in $t$-direction.\\
\noindent{\bf Numerical dispersion (Fermion doubling).} Numerical dispersion refers in the quantum relativistic and high energy physics literatures, and more specifically in gauge field theory as Fermion doubling \cite{cpc2012,potz1,potz2}. The latter occurs when the dispersion relation is not exactly satisfied, which generates artificial modes, and lead to chiral symmetry breaking \cite{cpc2012}. Standard real space methods usually generate these artifical fermionic states, while only one state exists at the continuous level; see  \cite{cpc2012} and associated references.  Mathematically those structures appear in the graph  of the eigenvalues of the amplification factor of the scheme (in Fourier-space). While the amplification factor of the continuous systems possesses 2 eigenvalues with conical graphes $\pm\sqrt{c^2|{\bf k}|^2+m^2c^4}$, real approximation methods have amplification factors (see \cite{strikwerda}) having eigenvalues with sinusoidal contribution, hence generated a periodic conical structures (in Fourier space). We refer the interested readers to \cite{potz2}, where this question is particularly well addressed for real space approximation. In order to discuss this question with the PINN-framework, we consider $\psi({\boldsymbol x},t)  \in \C^2$ solution to
\begin{eqnarray*}
{\tt i}\partial_t \psi -{\tt i}c{\boldsymbol \sigma}\cdot \nabla \psi + mc^2\sigma_z  \psi & = & {\bf 0} \, ,
\end{eqnarray*}
where we have denoted ${\boldsymbol \sigma}=(\sigma_x,\sigma_z)$. The Fourier transform in space is denoted $\widehat{\psi}({\boldsymbol k},t)$ with ${\boldsymbol k}=(k_x,k_y)^T$, and $\widehat{\psi}_0({\boldsymbol k})$ denotes the Fourier transform of the initial data. Formally, we get 
\begin{eqnarray*}
\widehat{\psi}({\boldsymbol k},t) & = & G({\boldsymbol k},t)\widehat{\psi}_0({\boldsymbol k}) \, ,
\end{eqnarray*}
where the $2\times 2$ (amplification) matrix reads
\begin{eqnarray*}
G({\boldsymbol k},t) & = & \exp\Big[-{\tt i}\big(ct{\boldsymbol \sigma}\cdot {\boldsymbol k} + mc^2\sigma_z \big)\Big] \, .
\end{eqnarray*}
The corresponding eigenvalues $h_{\pm}({\boldsymbol k}) = \exp\big(\mp {\tt i}t\sqrt{|{\boldsymbol k}|^2c^2+m^2c^4} \big)$ with absolute value equal $1$, corresponding to the conservation of mass (or $L^2$-norm). Rewriting $G=PDP^{-1}$, where $D=\textrm{diag}(h_{-},h_{+})$ and the transition matrix reads
\begin{eqnarray*}
P({\boldsymbol k}) = \left(
\begin{array}{cc}
 -\sqrt{|{\boldsymbol k}|^2c^2+m^2c^4}+mc^2 & c k_x - {\tt i}ck_y  \\
 c k_x + {\tt i}ck_y & \sqrt{|{\boldsymbol k}|^2c^2+m^2c^4}-mc^2 
\end{array}
\right) \, .
\end{eqnarray*}
We set $\widetilde{\psi}({\boldsymbol k},t) := P^{-1}({\boldsymbol k})\widehat{\psi}({\boldsymbol k},t)$ so that $\widetilde{\psi}=D\widetilde{\psi}_0$, and
\begin{eqnarray*}
\left.
\begin{array}{lcl}
\widetilde{\psi}_{1}({\boldsymbol k},t) & = & \exp\big(+{\tt i}t\sqrt{|{\boldsymbol k}|^2c^2+m^2c^4} \big) \widetilde{\psi}_{0;1}({\boldsymbol k}) \, ,\\
\widetilde{\psi}_{2}({\boldsymbol k},t) & = & \exp\big(-{\tt i}t\sqrt{|{\boldsymbol k}|^2c^2+m^2c^4} \big) \widetilde{\psi}_{0;2}({\boldsymbol k}) \, .
\end{array}
\right.
\end{eqnarray*}
We then define $\mu_{i}$ ($i=1,2$)
\begin{eqnarray}\label{mu}
\mu_{i}({\boldsymbol k,t}) = {\tt i}\log\Big[\cfrac{\widetilde{\psi}_i({\boldsymbol k},t)}{\widetilde{\psi}_{0;i}({\boldsymbol k},t)}\Big] = (-1)^{i}t\sqrt{|{\boldsymbol k}|^2c^2+m^2c^4}\, .
\end{eqnarray}
The graphs of  $\mu_{1,2}$ are reported in Fig. \ref{fig00}.

\begin{figure}[hbt!]
\begin{center}
\includegraphics[height=6cm,keepaspectratio]{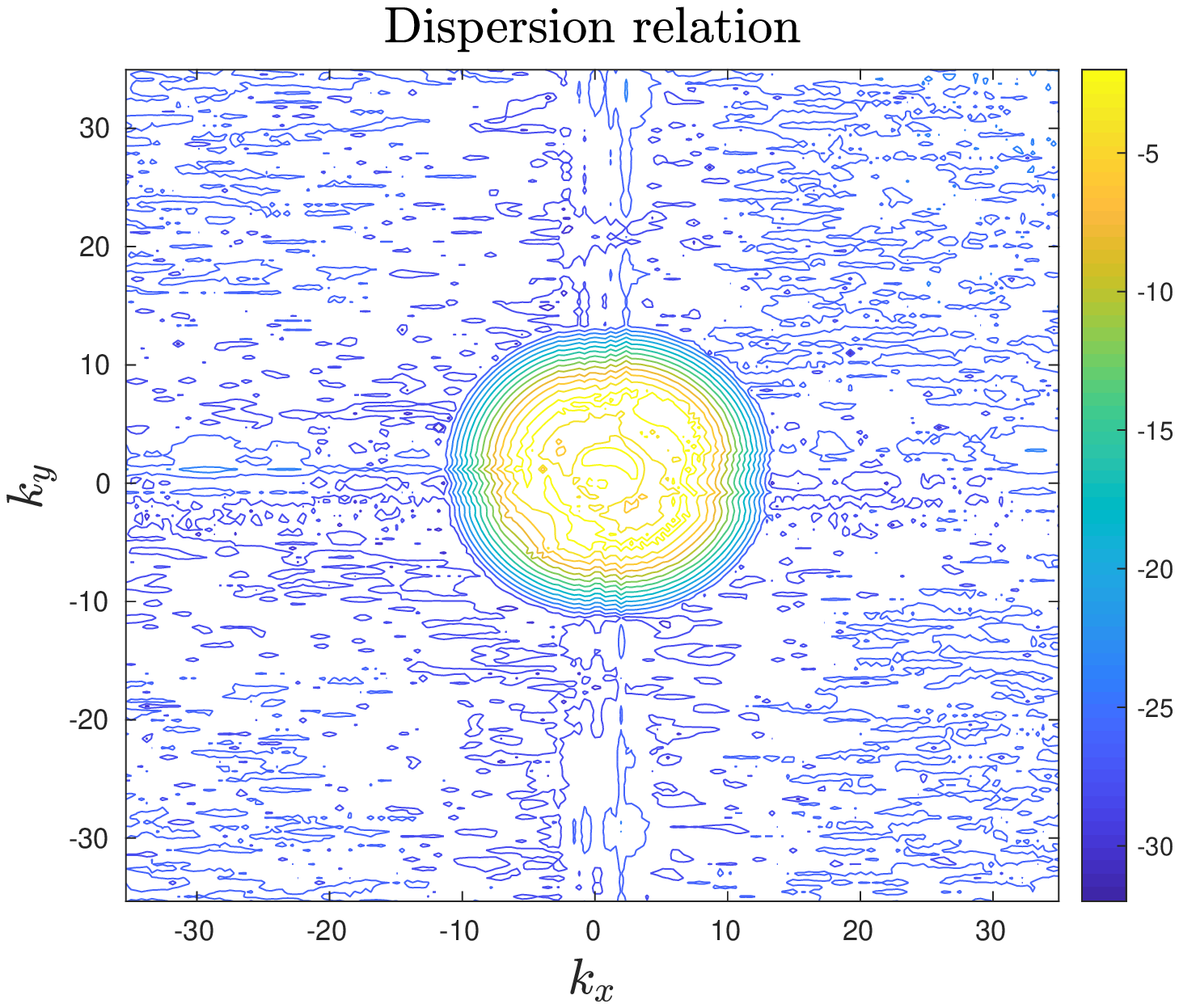}
\includegraphics[height=6cm,keepaspectratio]{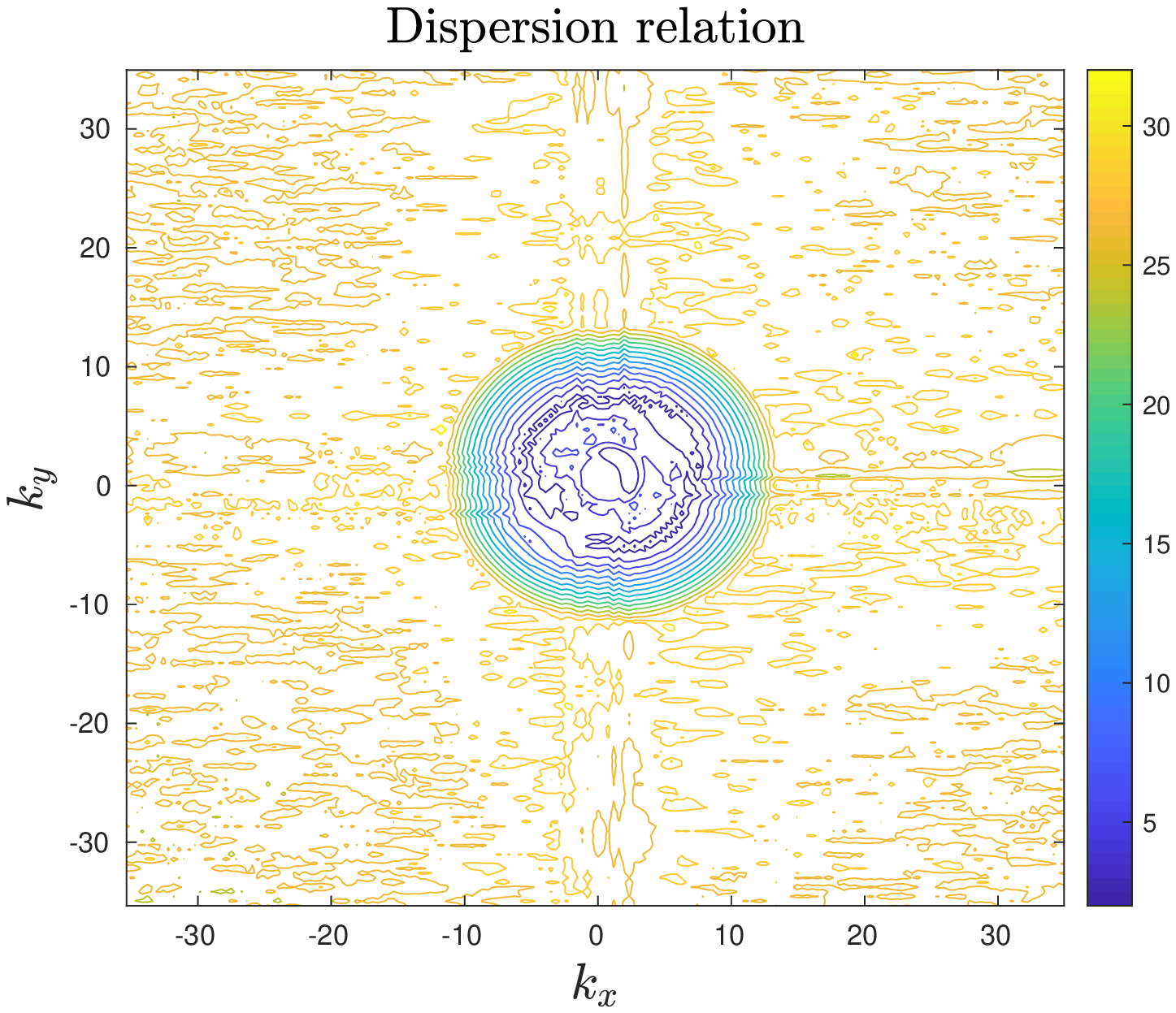}
\end{center}
\caption{{\bf Experiment 2.} Numerical dispersion relation \eqref{zeta}: graphs of  $\zeta_{1}$ (Left) and  $\zeta_2$ (Right).}
\label{fig0}
\end{figure}

The objective is hence to compare the above dispersion relation with the one obtained numerically thanks to the PINN algorithm. In this goal, we mimic the above computation in the PINN framework for a numerical solution ${\bf N}(\overline{{\boldsymbol p}},{\boldsymbol x},t) \in \R^4$. In order to properly reproduce the above calculations, we set ${\bf M}=(M_1,M_2):=(N_1+{\tt i}N_3,N_2+{\tt i}N_4)$ which takes it values in $\C^2$ and which is a PINN approximation to $\psi$, as described in Section \ref{sec:pinn}.  We denote by $\widehat{{\bf M}}$ the (discrete) Fourier transform in space of ${\bf M}$, and we set $\widetilde{{\bf M}}(\overline{{\boldsymbol p}},{\boldsymbol k},t) = P^{-1}({\boldsymbol k}){\bf M}(\overline{{\boldsymbol p}},{\boldsymbol k},t)$ so that $\widetilde{{\bf M}}=D\widetilde{{\bf M}}_0$. We then propose to compare $\mu_i$ with $\zeta_i$ for $i=1,2$, defined as follows
\begin{eqnarray}\label{zeta}
\zeta_{i}({\boldsymbol k},t) = {\tt i}\log\Big[\cfrac{\widetilde{{\bf M}}_i({\boldsymbol k},t)}{\widetilde{{\bf M}}_{0;i}({\boldsymbol k})}\Big] \, .
\end{eqnarray}
More specifically for fixed $t$, we will represent the graph of ${\boldsymbol k}\mapsto \zeta_i({\boldsymbol k},t)$. \\
{\bf Experiment 2.} We propose the following experiment with $m=1$, $c=1$, $V=0$. We report the graph of $\mu_{i}$ versus the one of $\zeta_i$ for $i=1,2$. The space-time domain is $(-8,8)\times[0,2]$ and we impose Dirichlet boundary condition. Initially we take
\begin{eqnarray}\label{init}
\phi_0({\boldsymbol x}) & = & \exp\big(-2\|{\boldsymbol x}\|^2_2+{\tt i}{\boldsymbol k}_0\cdot {\boldsymbol x}\big)(1,0)^T \, ,
\end{eqnarray}
where ${\boldsymbol k}_0=(1,1)^T$. We takes $1.5\times 10^4$ epochs (training iteration) and $5\times 10^4$ training points. The chosen network possesses $8$ layers and $50$ neurons. The converged solution is reached with a train loss of $9 \times 10^{-7}$ and test loss of $1.18\times 10^{-6}$. We report in Fig. \ref{fig0} the graph of $\zeta_1$ (Left) and $\zeta_2$ (Right) at $T=2$ obtained as described above. This example shows that the PINN-method is free from fermion doubling; the latter would indeed  correspond to the generation of artificial modes (typically in the form of periodic conical structures regularly spaced; see \cite{potz2}).\\

\begin{figure}[hbt!]
\begin{center}
\includegraphics[height=6cm,keepaspectratio]{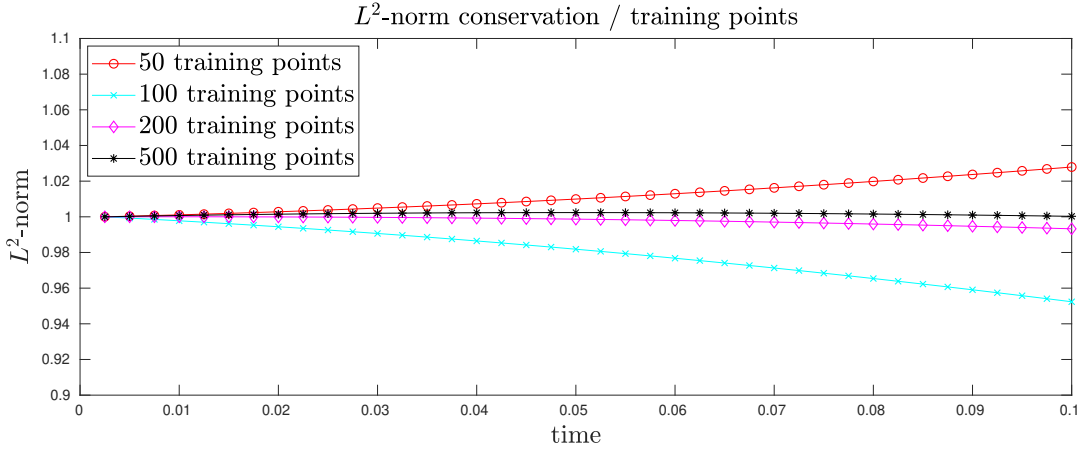}
\end{center}
\caption{{\bf Experiment 3.} $L^2$-norm conservation as function of time and with $50$, $100$, $200$, $500$ training points: $\{(t,\sum_{i=1}^4|N_i(\cdot,t)|_{L^{2}(\Omega);\R}^2), \, \, \, t \in (0,10^{-1})\}$.}
\label{figL2}
\end{figure}

\noindent{\bf $L^2-$norm conservation of the density for long physical times}. The preservation of the $L^2$-norm of the solution is a fundamental physical property, which can be interpreted as a mass conservation; that is, for any $t\geq 0$, we expect that the approximate solution satisfies $\|{\boldsymbol N}(\overline{{\boldsymbol p}},\cdot,t)\|_{L^2((\Omega);\R^4)} =1$. In order to preserve the $L^{2}$-norm, it is possible to directly impose a constraint within the loss function as described above. However, let us recall that in the above setting, the initial condition is imposed through a condition on the boundary $\Omega \times \{0\}$ at $t=T$. Hence, without imposing this additional constraint, only the boundary$/$initial condition ensures the PINN solution to the Dirac equation will not be null. As a consequence, for long physical times (hence requiring a very large number of training points, in particular in the time direction) the minimization of the loss function may lead to a solution with $L^2$-norm close to zero. This was already observed in \cite{grad}. We propose here to illustrate the time-domain decomposition which was proposed in Section \ref{sec:pinn}.\\
\noindent{\bf Experiment 3.}
In order to illustrate the principle of the time-domain decomposition approach, we just need to check if, on a sufficiently small time interval the norm of the density, can be properly preserved.  Indeed, at the end of the computation, the solution at final time can then be used as initial condition for the IBVP on the next time interval.  We propose the following illustration, with $t_1-t_0=10^{-1}$. As the time interval is small enough etc, we can naturally select a relatively small number of training points. We represent in this experiment, the $L^2$-norm of the density as a function of time $\{(t,\sum_{i=1}^4|N_i(\cdot,t)|_{L^{2}(\Omega);\R}^2), \, \, \, t \in (0,10^{-1})\}$.  The computational domain in space is $\Omega=(-5,5)$, and we use null Dirichlet boundary condition. We select a network with $4$ hidden layers and $30$ neurons, and the learning rate is taken equal to $10^{-3}$ and epoch size equal to $100$. The initial data is a Gaussian \eqref{init}. We report in Fig. \ref{figL2} the $L^2$-norm of the density as a function of time, when we successively when we use $50$, $100$, $200$ and $500$ training points in total. This shows that, at least taking a large enough number of training points the PINN algorithm allows for a good conservation of the overall $L^2$-norm. Thanks to the strategy developed in \eqref{DDT1}, \eqref{DDT2}, it is possible to ensure the norm conservation on very long physical times by decomposing the overall time-domain in small time-subdomains.\\

\begin{figure}[hbt!]
\begin{center}
\includegraphics[height=6cm,keepaspectratio]{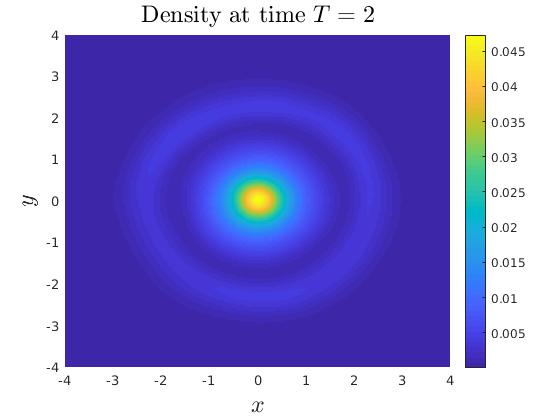}
\includegraphics[height=6cm,keepaspectratio]{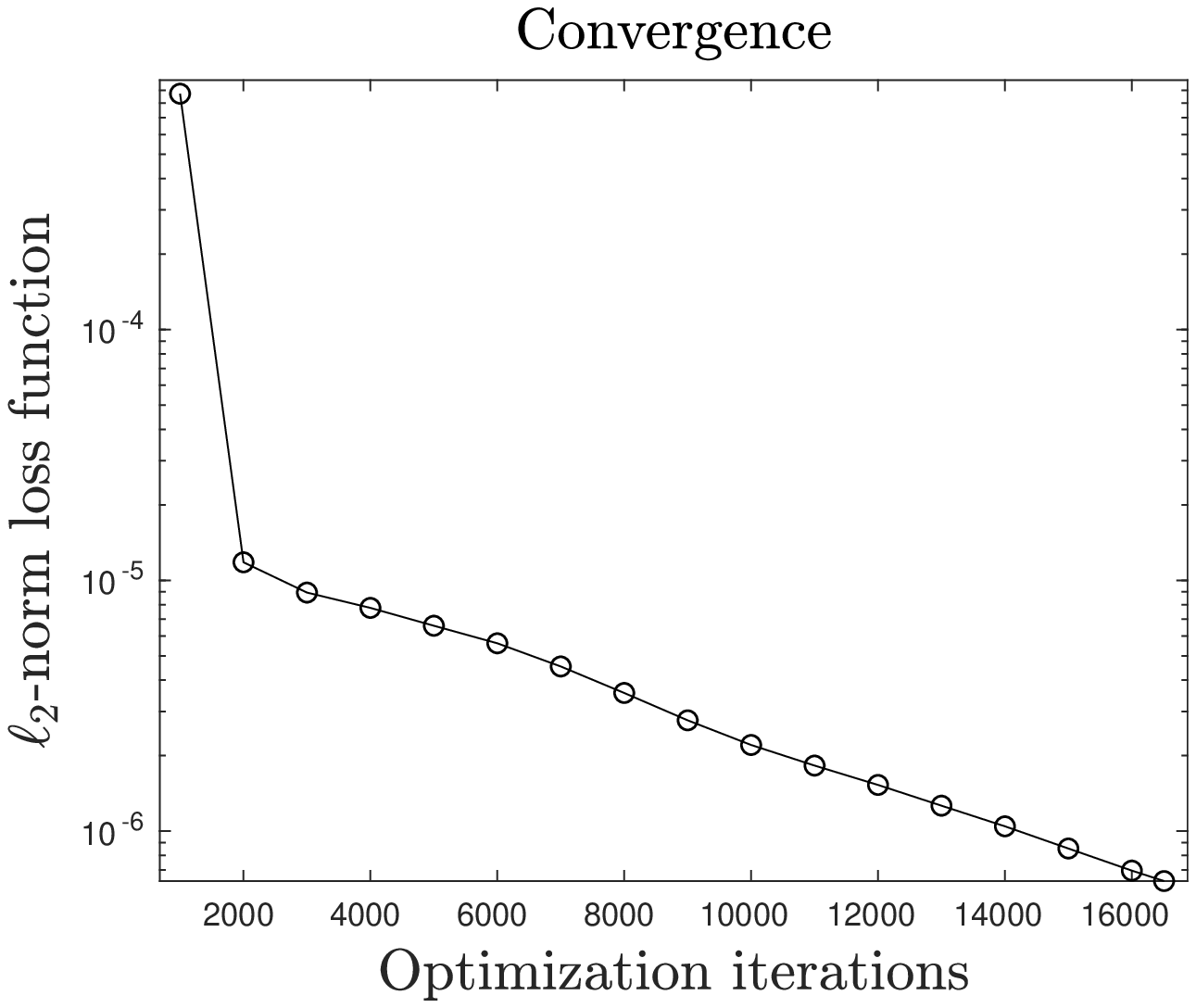}
\end{center}
\caption{{\bf Experiment 4.} (Left) Density at time $T=2$. (Right) Loss function convergence.}
\label{fig2}
\end{figure}

\begin{figure}[hbt!]
\begin{center}
\includegraphics[height=6cm,keepaspectratio]{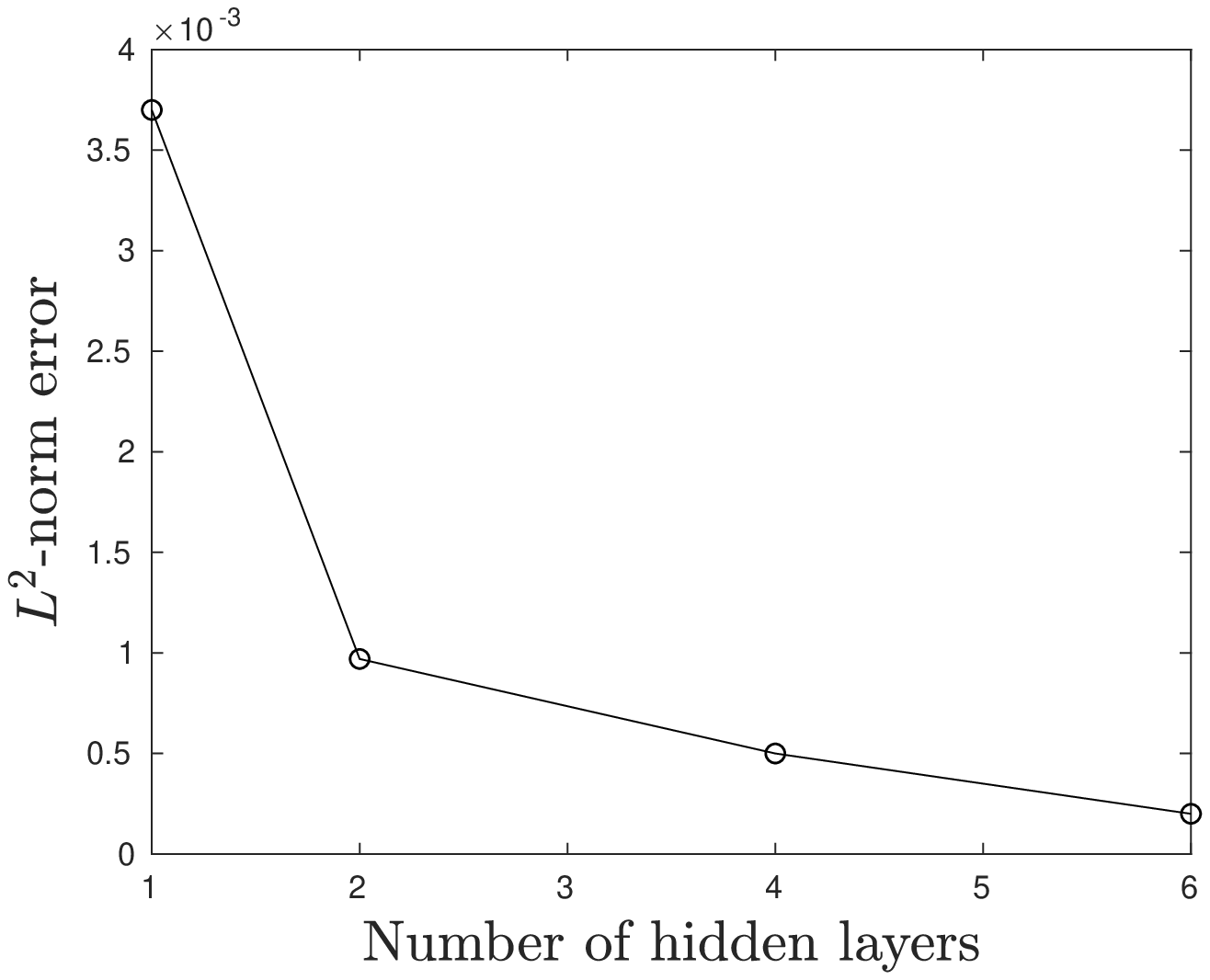}
\includegraphics[height=6cm,keepaspectratio]{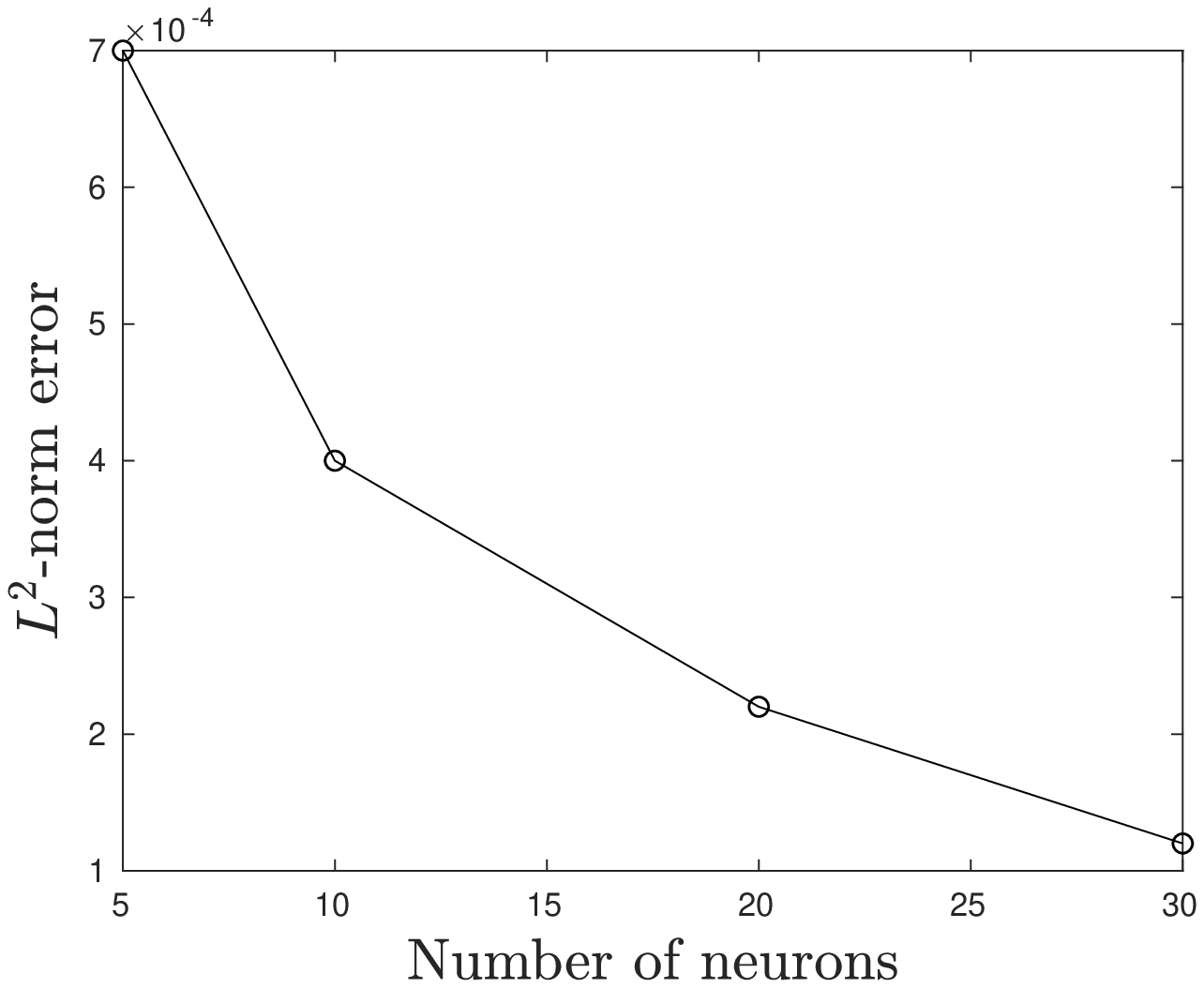}
\end{center}
\caption{{\bf Experiment 4.} $L^2$-norm ``error'' between density of reference at $T=2$. (Left) $1$, $2$, $4$, $6$ hidden layers and $40$ neurons (Right) $4$ hidden layers and $5$, $10$, $20$, $30$ neurons.}
\label{fig3}
\end{figure}

\noindent{\bf Computational error}. In the following experiment, we are interested in the behavior of the PINN algorithm when ``refining'' the PINNs; that is by increasing the number of hidden layers$/$neurons for a large and fixed number of training points. Let us notice that the following test does however not provide or illustrate overall convergence of the PINN algorithm. We refer to \cite{cv-pinn} for a first rigorous analysis of the PINN convergence. Notice however, that a full mathematical or numerical analysis of the PINN method is still to be achieved. \\
\noindent{\bf Experiment 4.} We consider the same setting as above, but on $(-4,4)^2\times [0,T]$, with ${\boldsymbol k}_0=(0,0)^T$ in \eqref{init}, and $V({\boldsymbol x})=-1/\sqrt{\|{\boldsymbol x}\|^2+\varepsilon^2}$ with $\varepsilon=0.25$. The objective of this experiment is to illustrate the behavior of the PINN algorithm when the NN is refined. In this goal, we report on Fig. \ref{fig2} (Left) the $L^{2}\big((-4,4)^2\big)$-error between the density of reference at $T=2$ computed with $8$ hidden layers and $50$ neurons, with i) the solutions with $4$ hidden layers and $5$, $10$, $20$, $30$ neurons, ii) the solutions with $1$, $2$, $4$, $6$ hidden layers and $40$ neurons. The minimization of the loss function is reported in Fig. \ref{fig2} (Right). The train (resp. test) loss value is $4.6\times 10^{-6}$ (resp. $3.4\times 10^{-6}$).  
We report in Fig. \ref{fig3} (Left), the $L^2$-norm between the density of reference at $T=2$ and the density computed using $1$, $2$, $4$, and $6$ hidden layers with $40$ neurons. In Fig. \ref{fig3} (Right.), we report the $L^2$-norm between the density of reference at $T=2$ and the density computed on $4$ hidden layers and $5$, $10$, $20$, $30$ neurons. Both tests illustrate the expected consistency of the overall PINN strategy applied to the TDDE.\\
\\
\subsection{Some physical experiments}
\begin{figure}[hbt!]
\begin{center}
\includegraphics[height=6cm,keepaspectratio]{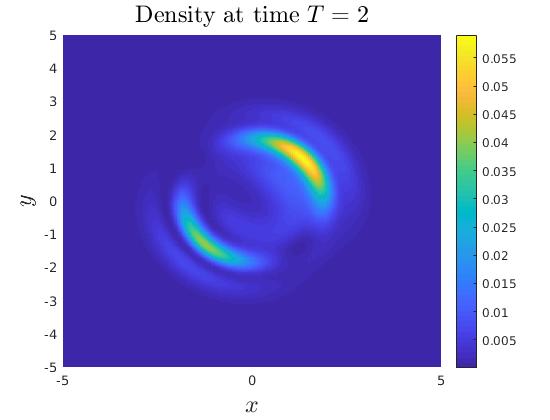}
\includegraphics[height=6cm,keepaspectratio]{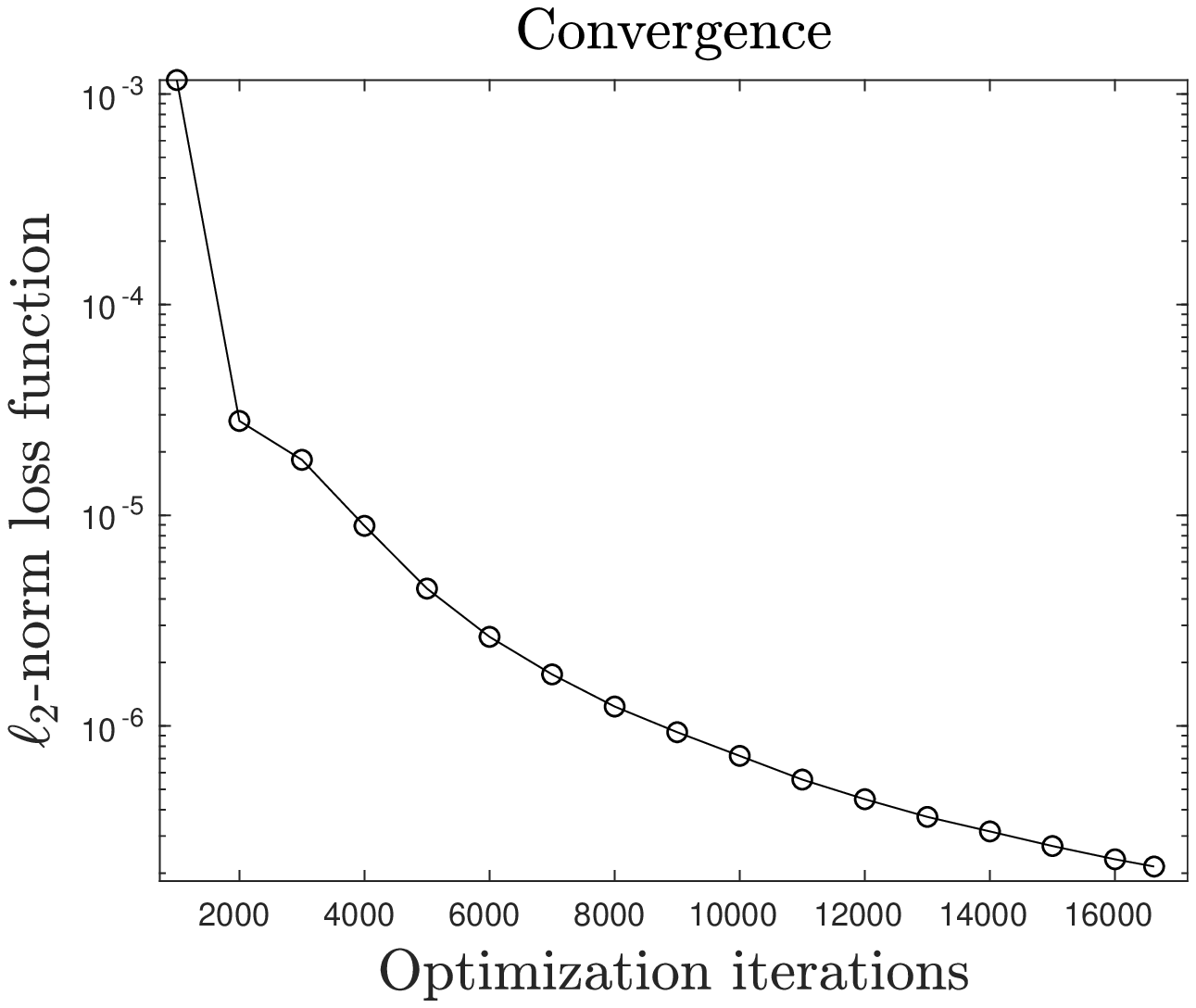}
\end{center}
\caption{{\bf Experiment 5.} (Left) Density at time $T=2$. (Right) Loss function convergence}
\label{fig1}
\end{figure}

\begin{figure}[hbt!]
\begin{center}
\includegraphics[height=4.25cm,keepaspectratio]{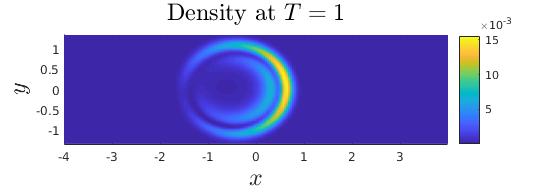}\\
\includegraphics[height=4.25cm,keepaspectratio]{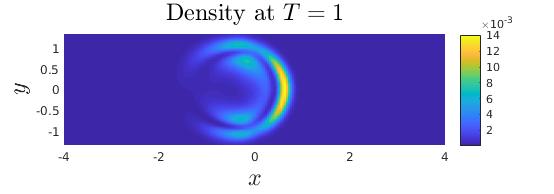}\\
\includegraphics[height=4.25cm,keepaspectratio]{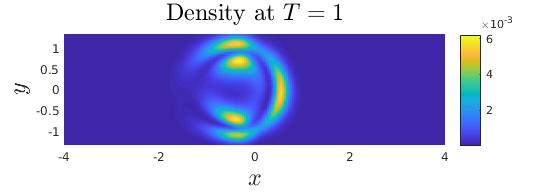}\\
\includegraphics[height=4.25cm,keepaspectratio]{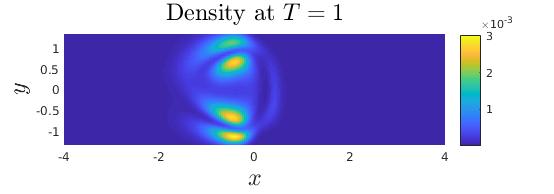}
\end{center}
\caption{{\bf Experiment 6.} From top to bottom: density at $T=2$ with i) $V_0=\sqrt{5}$, ii) $V_0=2\sqrt{5}$, iii) $V_0=3\sqrt{5}$, iv)  $4\sqrt{5}$.}
\label{fig4}
\end{figure}

\begin{figure}[hbt!]
\begin{center}
\includegraphics[height=4.25cm,keepaspectratio]{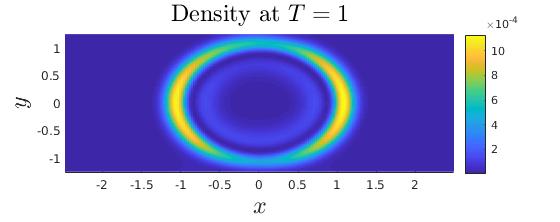}\\
\includegraphics[height=4.25cm,keepaspectratio]{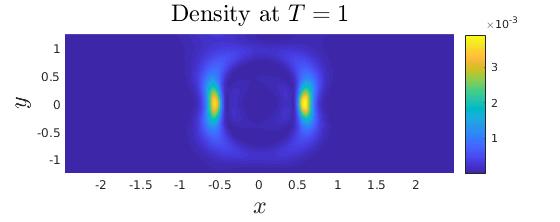}
\end{center}
\caption{{\bf Experiment 7.} (Top) Density at $T=1$ in flat space. (Bottom) Density at $T=1$ in curved space.}
\label{fig5}
\end{figure}

In this section we propose some more physics-oriented simulations to illustrate the PINN algorithm.\\
\noindent{\bf Experiment 5.} We consider the two-dimensional Dirac equation on the interval $(-5,5)^2\times [0,2]$. In this first experiment we take  $m=1$ and $V=0$. The initial condition is a wavepacket \eqref{init} where ${\boldsymbol k}_0=(1,1)^T$. The PINN is made of $3$ input layers, $8$ hidden layers with $40$ neurons and $4$ output layers. The number of training points is $10^4$ for the interior domain and boundary. The number space-time grid points is $75^2\times 20$ and $\tanh$ is chosen as the transfer function. The epoch size is taken equal to $10^3$ in the SGD with a learning rate of $10^{-4}$. In Fig. \ref{fig1}, we report the density ${\boldsymbol x}\mapsto \rho_N({\boldsymbol x},T)$ at time $T=2$ (Left). At convergence the mean residual (loss function value) is given by $1.12 \times 10^{-4}$. The train (resp. test) loss value is $2.2\times 10^{-7}$ (resp. $2.7\times 10^{-7}$). We report the loss function $\ell_2$-norm as function of optimizer iterations in Fig. \ref{fig1} (Right), illustrating the convergence of the minimization process.\\
\noindent{\bf Experiment 6.} The next experiment in connection with the {\it Klein paradox} \cite{cpc2012,KP}. The latter is a phenomenon which occurs in relativistic physics, and which corresponds to planewaves to cross a potential barrier $V_0$, when the energy $E$ of the incoming wave is lower than the potential. More specifically, wave transmission is possible for $E<V_0-mc^2$, thanks to the negative energy solution describing anti-fermions, while the incoming and reflected part parts are the electron wavefunction, see  \cite{cpc2012}. We consider the following potential
\begin{eqnarray*}
V(x,y) & = & \cfrac{V_0}{2}\Big(1+\tanh\Big(x/L\Big)\Big)\, ,
\end{eqnarray*}
where $L\ll 1$, and initial condition
\begin{eqnarray}\label{init2}
\phi_0({\boldsymbol x}) & = & \exp\big(-20\|{\boldsymbol x}-{\boldsymbol x}_0\|^2_2+{\tt i}{\boldsymbol k}_0\cdot {\boldsymbol x}\big)(1,C)^T \, ,
\end{eqnarray}
with ${\boldsymbol k}_0=(2,0)$, $C=k_{0;x}/(1+k_{0;x}^2+1)$ and ${\boldsymbol x}_0=(-1,0)^T$. The PINN algorithm is applied with $5$ hidden layers, $40$ neurons. We report in Fig. \ref{fig4} the density at $T=0$, then $T=1$ with $V_0=0\sqrt{5}$, $2\sqrt{5}$, and $3\sqrt{5}$, $4\sqrt{5}$. This tests shows that even a part of the wavefunction can still partially cross a potential barrier when $E<V_0-mc^2$ which refers as the Klein paradox.\\
\noindent{\bf Experiment 7.} This last experiment is about the modeling of electron  propogation on strained graphene surface. We here illustrate the fact that the implementation of the PINN algorithm is not modified by the ``complexification'' of the Dirac operator. Basically, we here have $m=0$, and we arbitrarily take $\rho$ has a Gaussian function loosely corresponding to a Gaussian surface \cite{pre}, and we neglect the pseudomagnetic field for simplicity \cite{pre3}. More specifically, we take
\begin{eqnarray*}
\rho({\boldsymbol x}) & = & 1 + \alpha\exp(-{\|{\boldsymbol x}-{\boldsymbol c}_{+}\|^2}) + \alpha\exp(-{\|{\boldsymbol x}-{\boldsymbol c}_{-}\|^2}) \, ,
\end{eqnarray*}
where $\alpha=5$, and ${\boldsymbol c}_{\pm}=(\pm 1,0)$. Initially, we consider a Gaussian initial wavefunction \eqref{init} with ${\boldsymbol k}_0=(0,0)^T$. We compare the evolution of the wavefunction in flat and curved spaces, see \eqref{fig5}. We observe the focusing effect of the graphene surface deformation, which is coherent with the computations performed in \cite{pre}. The changes in the Dirac equation (space-dependent coefficient) do not modify$/$complexify the PINN algorithm, but naturally modifies the complexity for minimizing the loss function.


\section{Conclusion}\label{sec:conclusion}
In this paper, we applied the celebrated PINN algorithm to solve the time-dependent Dirac equation in different frameworks, including relativistic quantum physics. Some experiments have been presented to illustrate some mathematical and physical properties when applied to the Dirac equation.  Due to its extreme simplicity and flexibility, this methodology is a very promising tool for simulating intense laser-molecule interactions in the relativistic regime, which will be the topic of a forthcoming paper. 

\bibliographystyle{unsrt}
\bibliography{refs}
\end{document}